\documentclass[
prl
,preprint%
 ,
,amssymb,floatfix,amsmath, aps,superscriptaddress,showpacs]{revtex4}
\usepackage{graphicx}
\usepackage{dcolumn}
\usepackage{bm}

\begin{document}

\title{Observation of vortex coalescence in the anisotropic
\\ spin-triplet superconductor Sr$_{2}$RuO$_{4}$}

\author{V.O. Dolocan}
\affiliation{CRTBT-CNRS, 25 Avenue des Martyrs, 38042 Grenoble,
France}
\author{C. Veauvy}
\affiliation{CRTBT-CNRS, 25 Avenue des Martyrs, 38042 Grenoble,
France}
\author{Y. Liu}
\affiliation{The Pennsylvania State University, University Park,
PA 16802,USA}
\author{F. Servant}
\affiliation{CRTBT-CNRS, 25 Avenue des Martyrs, 38042 Grenoble,
France}
\author{P. Lejay}
\affiliation{CRTBT-CNRS, 25 Avenue des Martyrs, 38042 Grenoble,
France}
\author{D. Mailly}
\affiliation{LPN-CNRS, Route de Nozay, 91460 Marcoussis, France}
\author{K. Hasselbach}%
\affiliation{CRTBT-CNRS, 25 Avenue des Martyrs, 38042 Grenoble,
France}
 \email{Second.Author@institution.edu}

\begin{abstract}
We present direct imaging of magnetic flux structures in the
anisotropic, spin-triplet superconductor Sr$_{2}$RuO$_{4}$ using a
scanning $\mu$SQUID microscope. Individual quantized vortices were seen
at low magnetic fields. Coalescing vortices forming flux domains were
revealed at intermediate fields. Based on our
observations we suggest that a mechanism intrinsic to the
material stabilizes the flux domains against the repulsive
vortex-vortex interaction. Topological defects like domain walls
can provide this, implying proof for unconventional chiral
superconductivity.
\end{abstract}

\date{\today}
\maketitle

\section*{}
A Type II superconductor allows flux penetration  in the form of quantized vortex
lines ($\phi_{0}$=h/2e), when placed in a
magnetic field greater than a material and sample-shape dependent
lower critical field H$_{c1}$. In an isotropic type II
superconductor the vortex lines have a shape of round cylinders
forming a triangular or square vortex lattice.
In the case of anisotropic superconductors vortex shape and vortex lattice structures depend on the symmetry and of the angle of the applied field.  Sr$_{2}$RuO$_{4}$ is a tetragonal, layered perovskite superconductor with a superconducting critical temperature (T$_{c}$) of
1.5 K\cite{Maeno}. The anisotropic superconducting properties of Sr$_{2}$RuO$_{4}$  are apparent in the penetration depth
anisotropy $\lambda_{c}$ = 3$\mu$m and $\lambda_{ab}$=0.15$\mu$m
 or in the anisotropy of the critical fields  $H_{c2}^{c}$ =0.075T and   $H_{c2}^{ab}$=1.5T\cite{MM} and are related to its two dimensional Fermi surface.

 In the past decade much of
the interest on Sr$_{2}$RuO$_{4}$ has derived from the theoretical
suggestion\cite{Rice} and subsequent experimental support\cite{MM,MRS}
that Sr$_{2}$RuO$_{4}$ is a spin-triplet, chiral Óp-waveÓ superconductor. 
 NMR measurements have observed that the spin
susceptibility is unchanged upon entering the superconducting
state\cite{Ishida},
NQR measurements reveal the absence of a Hebel-Slichter peak in
1/T$_{1}$T\cite{NQR}   and T$_{c}$ is strongly suppressed by non-magnetic
impurities\cite{Mackenzie}.  A spontaneous magnetic field has been detected in
the superconducting phase indicating the breaking of the time
reversal symmetry (TRS)\cite{Luke}. A TRS breaking state implies a
multiple component order parameter. The microscopic pairing mechanism is still
under debate.

Among the possible symmetries of the p-wave state the order parameter
$\mathbf{d}$($\mathbf{k}$)=$\widehat{z}$(k$_{x}$$\pm$ ik$_{y}$)\cite{SU} corresponds closest to the experimental results. The spin of the Cooper pairs lies in the
basal plane (equal spin pairing) with the $\mathbf{d}$ vector in the
c direction.
However this form of $\mathbf{d}$ usually gives a nodeless gap and seems
inconsistent with power law dependences observed experimentally in
many quantities as for e.g. the specific heat\cite{Nishizaki,Deguchi}. An interlayer
coupling\cite{Zit} (multi-band model)
was proposed to overcome this
dilemma along with an orbital dependent superconductivity\cite{ARS}.  The superconductivity originates from an active band $\gamma$ and is induced in the passive bands afterwards through inter-band interaction. The essential order parameter keeps the symmetry $\widehat{z}$(k$_{x}$$\pm$ ik$_{y}$) in the active band with an anisotropic gap.

Degenerate TRS breaking states can appear in the form of domains in
the superconducting state. Domain walls\cite{Sigrist} would separate
regions of degenerate order parameters with different surface
magnetization. These domains are predicted to act like fences
impeding the vortices in their movement. Thus  the visualization of vortices and the subsequent observation of
intrinsic pinning of vortices\cite{Dumont} are a necessary and important step
for resolving the unconventional superconductivity in
Sr$_{2}$RuO$_{4}$.

Small angle neutron scattering (SANS) measurements revealed
formation of a square vortex lattice\cite{Kealey} in
Sr$_{2}$RuO$_{4}$ after field cooling in fields ranging from 50 to 300 gauss applied along c-axis. The square
lattice and the detail of the magnetic field distribution around
the vortices were found to agree qualitatively with a two-component
p-wave Ginzburg-Landau theory\cite{Agterberg,Heeb,Kita}. However,
SANS is a bulk probe that is sensitive to the long-range
correlation in the vortex state rather than to the local
structure.
No scanning tunneling microscopy images of Sr$_{2}$RuO$_{4}$ succeeded to resolve vortices.
Here we present the first microscopic images of the magnetic
flux state in Sr$_{2}$RuO$_{4}$, using a custom-built $\mu$SQUID force microscope ($\mu$SFM) \cite{Veauvy}. The $\mu$SFM is a
sensitive tool for observing individual vortices on a local scale with a spatial resolution of 1$\mu$m. The $\mu$SQUID detects the
magnetic flux emerging perpendicularly from the sample's surface.
 
During the imaging, the
$\mu$SQUID moved in a plane above a cleaved ab surface of a single
crystal of Sr$_{2}$RuO$_{4}$. The
distance between the sample and the SQUID was kept constant at
1$\mu$m during scanning by a force detection scheme.
The crystal was grown by a floating
zone technique using an image furnace\cite{Servant}. Specific heat
measurements of crystals taken from the same single-crystal
rod showed volume superconductivity below a temperature of 1.31 K
and a transition width of less than 0.1K. 
 The sample has a plate
like shape with an estimated demagnetizing factor, N = 0.72.

\begin{figure}[t!]
  \includegraphics[width=6cm]{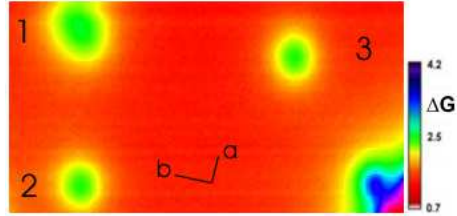}\\
  \caption{\label{Fig.1}A $\mu$SQUID force microscope ($\mu$SFM) image of the flux structure above
an ab face of Sr$_{2}$RuO$_{4}$, obtained at T = 0.36K, H = 0.1G (H
$\parallel$ c, field cooling at 0.1 G). The panel corresponds to an
imaging size of 31$\mu$m $\times$ 17$\mu$m. The color legend on the
right indicates values of the flux density in gauss. The orientation
of the crystallographic a-b axes is shown.}
\end{figure}

 Round flux structures are seen after cooling the crystal
in a magnetic field of 0.1G applied along the c-axis,
(Fig.~\ref{Fig.1}). Integrating the magnetic field at locations 2 and 3 yields
$\sim$1$\phi_{0}$ of flux while the
integration at location 1 corresponds to $\sim$2$\phi_{0}$.
The measured field profile at locations 2 and 3 can be well adjusted to
 the model \cite{Kirtley} of a single quantized
vortex using values for the  scanning height between 1-2 $\mu$m  and a penetration depth
$\lambda_{ab}$=0.15-0.2 $\mu$m.
The value for the in-plane penetration depth is in agreement with literature values.
The quantized amount of flux suggests the presence of
single vortices at locations 2 and 3, and a vortex pair at
location 1. The flux structure on the bottom right corner, which
is not quantized, is most likely due to the presence of a defect
at that location.  As the sample temperature decreases below T$_{c}$ and under applied
magnetic field vortices form in a superconductor 
and may stay pinned at fields lower than H$_{c1}$. The observed flux structures were seen to disappear
completely above T = T$_{c}$ = (1.35$\pm$0.05)K, in agreement with
the T$_{c}$ value determined previously in specific heat
measurements.

\begin{figure}[!]
  \includegraphics[width=9cm]{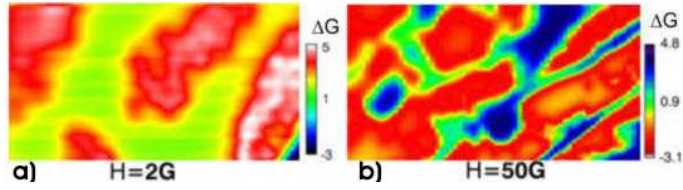}\\
  \caption{\label{Fig.2}$\mu$SQUID imaging of superconducting domain formation in  Sr$_{2}$RuO$_{4}$  for magnetic fields
applied perpendicular to the ab plane. The data of Fig.~\ref{Fig.2}a)  is acquired after field
cooling in 2 gauss,  the imaging area is 31$\mu$m $\times$
17$\mu$m and flux free domains appear green.
Fig.~\ref{Fig.2}b) shows the magnetic state after zero field cooling the sample
and applying then 50 gauss, the area is 62$\mu$m $\times$ 33$\mu$m. The measurement temperature is 0.35K.}
\end{figure}

After field cooling
the sample down to 0.35 K in an applied field of 2 gauss,  we
observed the presence of flux domains (Fig.~\ref{Fig.2}a).
The difference in flux density between the red (normal) and the
green (superconducting) regions is 3 gauss. Integration of the
flux pattern gives an average field of 1.4$\pm$0.2 gauss, close to
the applied field of 2 gauss. All the flux is condensed in
domains, leaving entire superconducting regions empty. These
domains are oriented 45$^{0}$ from the crystallographic axis. This orientation is in agreement with the flux line lattice orientation observed in SANS and $\mu$SR. Flux
domain structures appear also after zero field cooling (ZFC) the sample and applying 50 gauss at 0.35K. At equilibrium 5000 vortices
should be present at this field and vortices would
overlap so much that our SQUID would not resolve them, at best, we
expect to detect a weak modulation of the measured flux density.
We do observe magnetic field variation as large as 8 gauss between neighboring domains,
Fig.~\ref{Fig.2}b. 

\begin{figure}[!]
  \includegraphics[width=9cm]{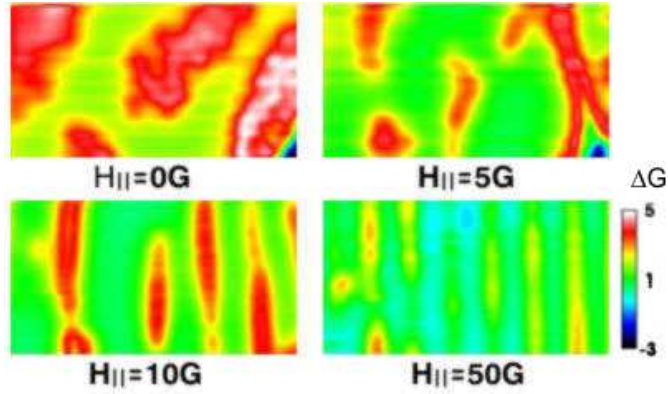}\\
  \caption{\label{Fig.3}$\mu$SM images of flux domains in
Sr$_{2}$RuO$_{4}$ at T = 0.36K after field cooling at various fields
as indicated. The imaging area is 31$\mu$m $\times$ 17$\mu$m. In all
cases, the field amplitude along the c-axis (H$_{\perp}$) was kept constant at 2G while the
in-plane field (H$_{ab}$) was set as indicated. The first panel is identical with the first one in Fig.~\ref{Fig.2}.
Field scales
in gauss are shown on the right; blue and green regions are
flux free regions.}
\end{figure}

 Vortices in Sr$_{2}$RuO$_{4}$ seem to attract each other and form
domains of magnetic flux. A complete
collapse of the vortices into one single domain is not observed,
probably due to the presence of weak pinning in the material.
Domain walls delimiting regions with different order parameters
(k$_{x}$+ik$_{y}$ and k$_{x}$-ik$_{y}$ ) could provide the scenario for weak intrinsic
pinning. Domain walls\cite{Sigrist} act as pinning regions for the vortices due
to the locally diminished condensation energy at the walls. Vortices are fenced
in by these walls at low magnetic fields and form domains of
magnetic flux.
In order to examine the stability of the domain configuration
the in-plane field was
raised while the c-axis field was kept at 2G. Fig.~\ref{Fig.3}
shows for increasing in-plane
fields how the condensed vortex structures rearrange freely in order to
accommodate the experimental conditions: At 5 gauss in-plane applied field the flux domains become slimmer and  above 10 gauss the flux
domains are seen to evolve into line-shaped structures. The
number of the flux lines was found to increase with the in-plane
field in a regular fashion. This regular increase of the flux
domain density and their temperature evolution (data not shown)
suggest that the flux structures are unrelated to any structural
defects in the crystal. Defect pinning\cite{Huse} of vortices
would interfere with regularly spaced vortex pattern (see also
ref.~\onlinecite{Poole} and references therein).

\begin{figure*}[!]
  \includegraphics[width=11cm]{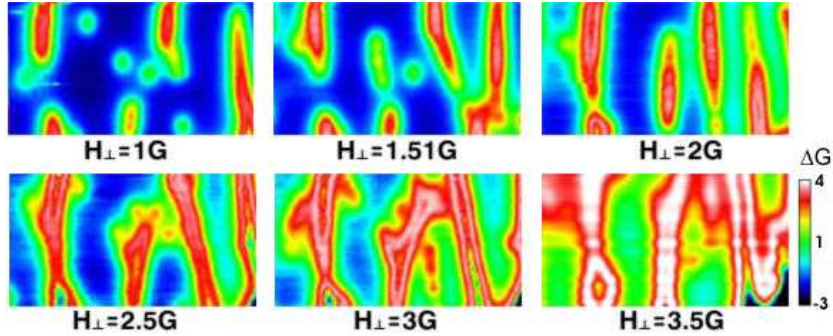}\\
  \caption{\label{Fig.4}Magnetic images of the flux structures in
Sr$_{2}$RuO$_{4}$ after field cooling keeping H$_{ab}$ constant at 10G and H$_{\perp}$ is varied
as indicated. The imaging area is 31$\mu$m $\times$ 17$\mu$m and the temperature is 0.36K. The flux density scale is shown on the right.}
\end{figure*}

 The line-shape flux structures resemble vortex chains observed in
vortex imaging experiments of YBa$_{2}$Cu$_{3}$O$_{7+\delta}$\cite{Gammel},
Bi$_{2}$Sr$_{2}$CaCu$_{2}$O$_{8+\delta}$\cite{Bolle} and
NbSe$_{2}$\cite{Hess}. Vortex chains appear when the applied field is close to the
in-plane direction of the anisotropic superconductor. Sr$_{2}$RuO$_{4}$
has an effective mass anisotropy ($\Gamma$) 40 times higher than NbSe$_{2}$
and 8 times higher YBa$_{2}$Cu$_{3}$O$_{7+\delta}$ but lower than
Bi$_{2}$Sr$_{2}$CaCu$_{2}$O$_{8+\delta}$. Consequently the
arrangement of the domains in lines may be driven by the
anisotropy, orienting the attractive interaction\cite{Grishin,Buzdin,Dolocan} between vortices
along the plane spanned by the anisotropy axis and the in-plane
applied field. Nevertheless we observe this coalescence of vortices
even when the in-plane field is absent which suggests that another
interaction mechanism is present.

How will the flux domain structure evolve as the in-plane field is further tilted toward the anisotropy axis?  In Figure~\ref{Fig.4}, the parallel field, H$_{ab}$, was
constant at 10 gauss and the perpendicular component was increased. Each data set was acquired after field cooling. In the first panel, flux domains and
individual vortices are clearly seen. As the perpendicular field
was increased more magnetic flux came out of the sample surface.
The flux domains attract the individual vortices and at higher
fields we see only domains of flux. The domains stretch in the
direction of the tilted field. Continuous flux domains are forming
for $H_{\perp}$ about 2 gauss. With increasing the field in
perpendicular direction the domains are starting to deform and
branch, reminiscent of the pattern formation in liquid crystals.
For perpendicular fields higher than 2 gauss, broad flux
structures split into two very narrow lines staying in close
proximity. These lines are narrower than individual vortices.

Vortex coalescence overcoming the usual vortex-vortex
repulsion\cite{Mohamed} is predicted for superconductors with
$\kappa$ close to $1/\sqrt{2}$, the
Ginzburg-Landau parameter $\kappa$ is defined as the ratio between
the penetration depth and the coherence length.
Superconductors with $\kappa<1/\sqrt{2}$ (type I) don't enter the vortex state but present, in
the case of flux penetration, the intermediate state, consisting
of meandering domains of normal and superconducting regions. A thin film of a type I superconductor may contain even vortices \cite{Huebner}.
An interesting limiting case constitutes Niobium as it is at the border between type I and type II superconductivity. It undergoes a first-oder transition at $H_{c1}$  accompanied by a magnetization jump $B_{0}$ and consequently presents\cite{Kerchner} an intermediate-mixed state if the demagnetizing factor N$\neq$0. This state is characterized by
the simultaneous presence of flux free regions and regions
containing a well-established vortex lattice with a lattice constant corresponding to $B{_0}$.
This state exists in a field range between H$^{*}$= $H_{c1}$(1-N) and $H_{c1}$(1-N)+$B_{0}$N.
Below is the Meissner phase, characterized by the absence of vortices, and above the Shubnikov phase with its vortex lattice. In the intermediate-mixed state an influence of
the crystal lattice on the orientation of the flux structures is
observed\cite{Huebner}.
Generally superconductors with $\kappa>1/\sqrt{2}$ (type II) have
always displayed vortex repulsion and a single quantized vortex
state.

In our experimental situation Sr$_{2}$RuO$_{4}$ has a $\kappa$ value $\sim$2 as the magnetic field is directed along the c-axis, a value significantly higher
than $\kappa$ of Nb. No first order transition in the magnetization curves of Sr$_{2}$RuO$_{4}$ is reported. Individual vortices, the signature of type II
superconductivity, are present in the sample at low fields after field cooling, but no domains are observed, clearly designating Sr$_{2}$RuO$_{4}$ as a type II superconductor.
At intermediate fields the vortices coalesce and leave only a few
individual vortices, no vortex lattice is observed. Above H$_{c1}$
(35 gauss, estimation based on $\mu$SQUID measurements) the domains persist, when the vortex lattice
formation should have set in. These experimental findings show that the
reasons for vortex coalescence in Sr$_{2}$RuO$_{4}$ are different
from those in conventional low $\kappa$ superconductors.

The remarkable flux patterns and its systematic variation with the
strength and the orientation of the field observed in
Sr$_{2}$RuO$_{4}$ can be related solely to intrinsic physical
processes in the superconducting state of this material. The
coalescence of vortices and splitting of flux domains into very
narrow lines are unique features in a type II
superconductor and may be due to the presence of domain walls separating regions of different
chirality of the order parameter. These domain walls would act as
corals containing magnetic flux. Furthermore the high mass anisotropy of
Sr$_{2}$RuO$_{4}$ will tend to stabilize vortex domains and
contributes to regular vortex pattern formation more reminiscent of liquid
crystals. 

\begin{acknowledgments}
We acknowledge the support of CNRS, and fruitful discussions with
V. Mineev, M. Zhitomirsky, M. Sigrist, G. Blatter, V.B.
Geshkenbein, and J. Flouquet. Y.L. is supported in part by US ONR.
\end{acknowledgments}

\end{document}